\documentclass[pdflatex,sn-mathphys-num]{sn-jnl}
\usepackage{graphicx,amsmath,amssymb,algorithm,algpseudocode}
\begin{document}
\title{A Butterfly-Inspired Agent Learns to Climb with Self-Generated Wingbeats}
\author[1,2]{ \fnm{Yixin} \sur{Chen}}\email{chenyixin@imech.ac.cn}

\author[1,2]{ \fnm{Jianfeng} \sur{Lin}}

\author*[1,2]{ \fnm{Shizhao} \sur{Wang}}\email{wangsz@lnm.imech.ac.cn}

\affil[1]{\orgdiv{State Key Laboratory of Nonlinear Mechanics}, \orgname{Institute of Mechanics}, \orgaddress{  \postcode{100190}, \state{Beijing}, \country{China}}}

\affil[2]{\orgdiv{School of Engineering Sciences}, \orgname{University of Chinese Academy of Sciences}, \orgaddress{ \postcode{101408}, \state{Beijing}, \country{China}}}

\abstract{Butterfly free climbing involves dynamic coupling between wing motion and unsteady aerodynamic forces that drive the animal’s ascent. Four-wing coordination determines how these forces combine to support weight and direct propulsion. How aerodynamic feedback shapes such coordination remains difficult to establish when wing kinematics are prescribed in experiments and simulations. Here, we construct a virtual flight environment with rapid aerodynamic feedback and train a butterfly-inspired agent using the soft actor–critic algorithm. Without a wingbeat generator, the agent spontaneously develops non-sinusoidal wingbeats and achieves reward-directed climbing. The climbing angle varies primarily with wingbeat frequency and mean angle of attack, and force-balance analysis links these adaptations to gravity projections along and normal to the flight direction. The aerodynamic forces are faithfully recovered only when the agent-generated wing motions are reconstructed with higher-order harmonics, rather than the commonly used first harmonic. Independent and combined modulation of frequency and mean angle of attack further reveals their nonlinear coupling in regulating climbing direction. These findings reveal how coordinated propulsion strategies emerge under physical constraints and provide a theoretical framework for studying strategy formation through interactions between organisms and their fluid environment.}

\keywords{biolocomotion, reinforcement learning, unsteady aerodynamics}

\maketitle

\section{Introduction}\label{sec1}

Flapping-wing animals perform aerial maneuvers during takeoff~\cite{johansson2021butterflies}, pursuit~\cite{mischiati2015internal}, and escape~\cite{muijres2014flies}. Free climbing couples wing-generated unsteady aerodynamic forces with center-of-mass translation under gravity~\cite{wang2016insect}. During climbing, these forces provide both weight support and the acceleration changing the flight velocity. Butterflies are representative four-winged insects whose hindwings contribute substantially to evasive maneuverability~\cite{jantzen2008hindwings}. How butterflies coordinate their wing motions to achieve different climbing directions remains unclear~\cite{le2019effects,le2019adaptive}. Dynamical models of monarch butterflies have examined hovering and forward climbing with prescribed periodic motions~\cite{tejaswi2021abdomen}. Identifying the physical principles of butterfly free climbing can establish a link between aerodynamic mechanisms and flight strategies in complex propulsion systems.

Experiments and high-fidelity simulations provide complementary approaches to studying biological propulsion. Motion reconstruction quantifies wing kinematics and flight velocity in free flight~\cite{kang2018experimental,bode2020reverse, lou2024wing}. Synchronized force and kinematic measurements in birds further show how lift and drag contribute to weight support, acceleration, and braking~\cite{chin2019birds}. Wing kinematics and flight velocity vary together during free flight, so their observed covariation alone does not isolate the aerodynamic contribution of each motion parameter. Controlled experiments and simulations allow selected kinematic parameters to be varied while others are held fixed.
Such comparisons reveal how specific kinematic adjustments change the balance between lift and thrust and the aerodynamic cost of generating these forces~\cite{noda2023interplay,lian2014characterization,johansson2024aerodynamic}. These studies quantify the aerodynamic consequences of selected wing motions. Prescribing wing kinematics breaks the feedback loop through which changes in flight velocity alter aerodynamic forces and, in turn, influence subsequent wing motion. Such comparisons therefore cannot reveal how wingbeat strategies emerge from the coupling between wing motion and flight velocity through unsteady aerodynamic forces.

Reinforcement learning offers a route to study how propulsion strategies emerge through interaction with a fluid environment~\cite{recht2019tour}. It has enabled agents to exploit vortices for efficient swimming, learn gliding strategies, and navigate atmospheric thermals~\cite{verma2018efficient,novati2019controlled,reddy2018glider}. A two-dimensional flapping wing can also learn non-harmonic, quasi-periodic motions from flow feedback~\cite{bi2025mimic}. In biomechanically detailed agents, however, coordination is often guided by reference motions or prescribed pattern generators rather than learned from physical feedback alone~\cite{aldarondo2024virtual,vaxenburg2025wholebody}. These motion priors support successful locomotion but make it difficult to distinguish coordination learned through interaction from that supplied in advance. Reinforcement learning in a virtual aerodynamic environment without prescribed wingbeat patterns therefore provides a way to investigate how aerodynamic feedback shapes butterfly climbing strategies.

Here, we develop a reinforcement-learning framework with an interactive aerodynamic environment to investigate the free-climbing strategies of a butterfly-inspired agent. Using direct numerical simulation data, we construct a quasi-steady model whose aerodynamic coefficients are predicted by a neural network and couple it with the agent's translational dynamics to form an interactive environment. A soft actor--critic agent~\cite{haarnoja2018soft} directly controls the angular jerks of four kinematic angles, allowing it to control its wings freely without prescribed wingbeats. By varying the reward weights for forward and vertical displacements, we obtain flight policies across different climbing angles and analyze the kinematic and aerodynamic characteristics of the free-climbing strategies formed by the agent.

\section{Model and methods}\label{sec2}

\subsection{Butterfly-inspired agent}\label{sec2.1}

\begin{figure}
\centering
\includegraphics[width=0.9\textwidth]{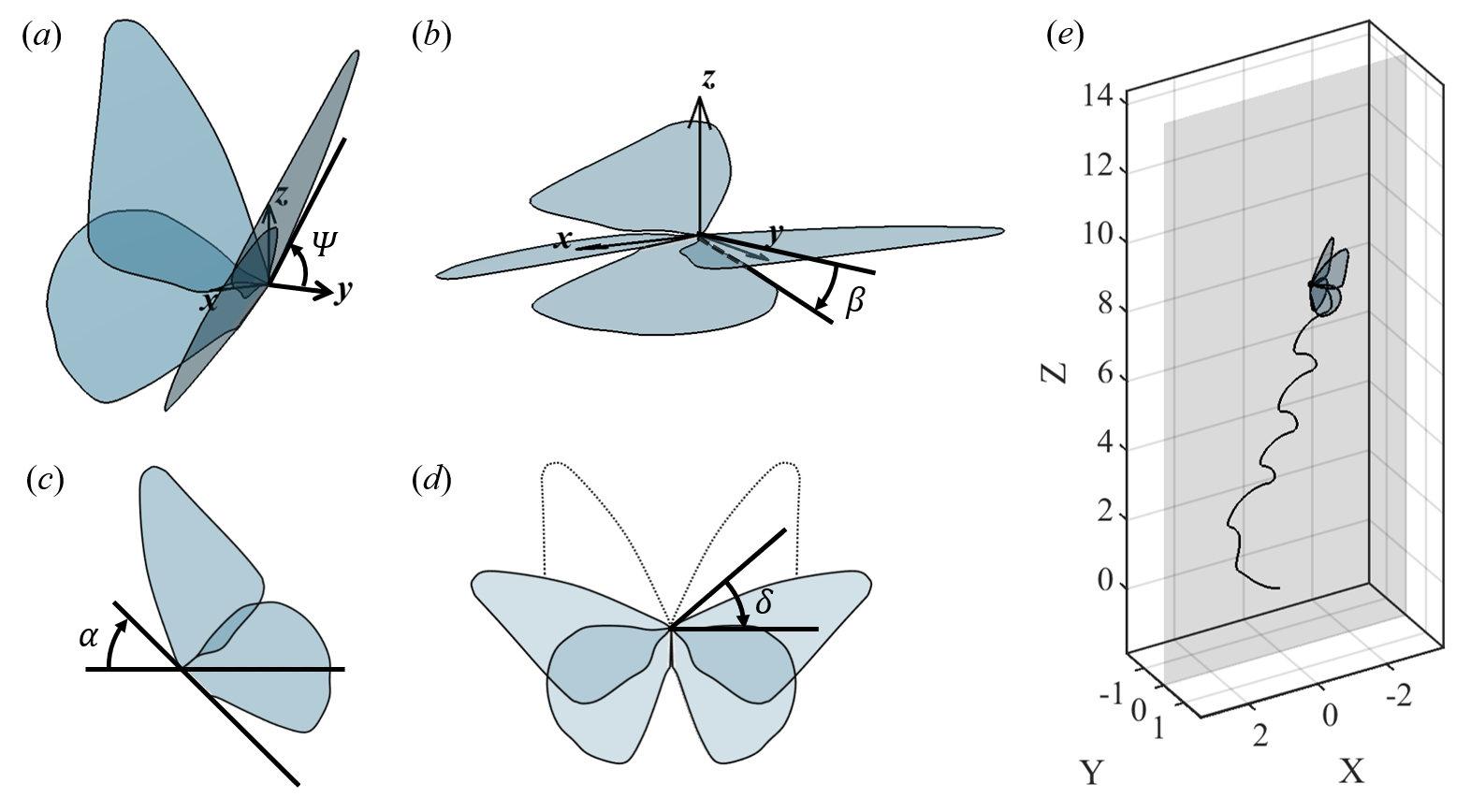}
\caption{Degrees of freedom and angular definitions of the butterfly-inspired agent. (a–d) Schematic definitions of the elevation angle ($\Psi$), slot angle ($\beta$), geometric angle of attack ($\alpha$), and sweep angle ($\delta$), respectively. The arrows indicate the positive direction of each angle. (e) The center-of-mass motion of the three-dimensional agent is constrained to the $x-z$ plane at $y=0$ (gray plane), with no lateral translation.}\label{figmodel}
\end{figure}

The agent comprises a pair of forewings and a pair of hindwings, with simplified geometry based on Ancel et al.~\cite{ancel2017aerodynamic}. Wing motion is described by four angles: elevation $\Psi$, geometric angle of attack $\alpha$, forewing--hindwing slot angle $\beta$, and sweep angle $\delta$. Their definitions and positive directions are shown in Fig.~\ref{figmodel}(a--d). The slot angle is the dihedral angle between the forewing and hindwing planes~\cite{chen2025streamwise}. The center of mass translates within the $x$--$z$ plane at $y=0$, while the wings retain three-dimensional motion (Fig.~\ref{figmodel}(e)).

The agent is trained using the soft actor--critic (SAC) algorithm. Its actions are the angular jerks of the four wing angles,

\begin{equation}
    \mathbf{a}_t
    =
    \dddot{\mathbf{q}}_t
    =
    \begin{bmatrix}
        \dddot{\Psi} &
        \dddot{\alpha} &
        \dddot{\beta} &
        \dddot{\delta}
    \end{bmatrix}^{\mathrm T}_t.
\end{equation}

Time integration converts these actions into angular accelerations, angular velocities, and angles. This formulation updates angular acceleration incrementally rather than assigning it independently at each step. No baseline waveform or wingbeat generator is imposed. The agent therefore controls the timing of stroke reversals and the shape of each wingbeat through interaction with the aerodynamic environment.

\subsection{Virtual physical environment}\label{sec2.2}

\begin{figure}
\centering
\includegraphics[width=0.9\textwidth]{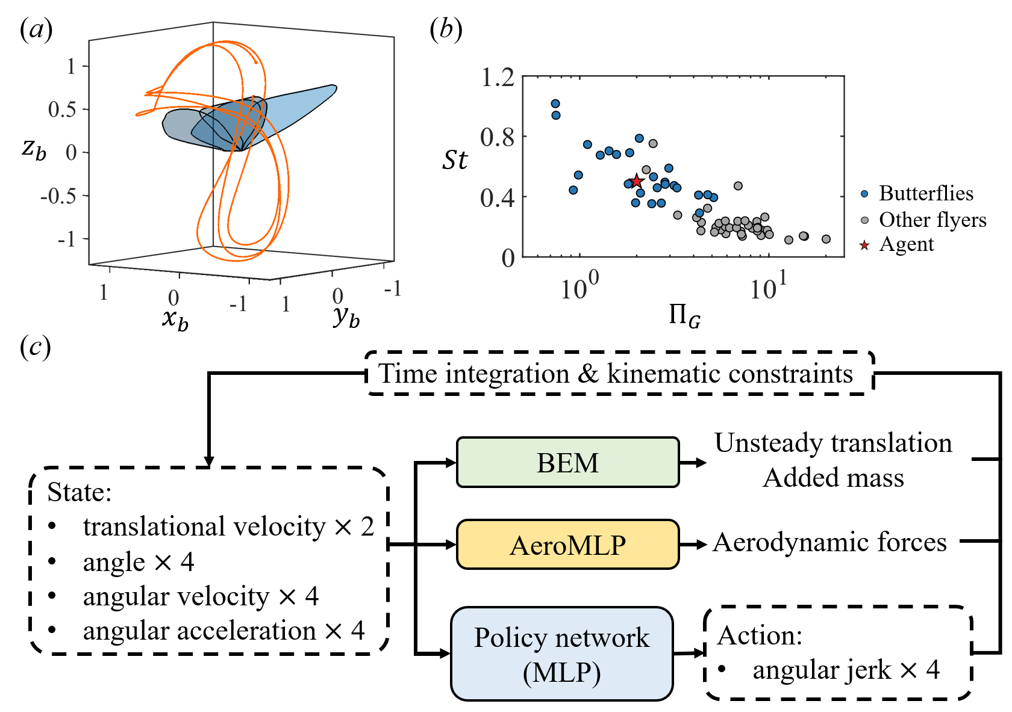}
\caption{Control framework of the butterfly-inspired agent. (a) Body-frame posture and wingtip trajectories generated through free control of the four kinematic angles. (b) Dynamic-pressure-to-weight ratio $\Pi_G$ and Strouhal number $St$ of flying animals~\cite{lou2024wing,sun2003lift,shyy2008aerodynamics,pennycuick1990predicting,pennycuick1996wingbeat,dudley1990biomechanics,ha2013relationship,wakeling1997dragonfly,liebe2007flow,fei2016importance,kang2018characterization,sridhar2016instantaneous,chen2022practical}, with butterflies highlighted in blue and the present agent marked by the red star. (c) Closed-loop interaction between the policy network and the virtual physical environment.}\label{fig2}
\end{figure}

The virtual environment couples aerodynamic prediction with translational dynamics and constrained time integration. Independent control of the four wing angles permits complex three-dimensional wingtip trajectories, as illustrated in Fig.~\ref{fig2}(a). Gravity is included explicitly to represent the force requirements of climbing. We assess the agent's operating regime using the dynamic-pressure-to-weight ratio $\Pi_G$ and the Strouhal number $St$, which characterize aerodynamic force relative to body weight and flapping motion relative to forward flight, respectively. The representative operating point lies within the range occupied by real butterflies in Fig.~\ref{fig2}(b). This comparison establishes similarity in these two nondimensional measures, rather than complete dynamical equivalence.

Figure~\ref{fig2}(c) summarizes the interaction loop. The policy receives the two translational velocities and the four wing angles, angular velocities, and angular accelerations, and outputs four angular jerks. Aerodynamic forces are evaluated by a multilayer perceptron for unsteady aerodynamic-coefficient prediction (AeroMLP). Trained against direct numerical simulation (DNS) forces, AeroMLP predicts state-dependent coefficients for a quasi-steady formulation based on Cai et al.~\cite{cai2021cfd}, with additional viscous-friction and clap-and-fling contributions. The flight state is first expressed in an equivalent level-flight frame using the flight speed and effective angle of attack. The predicted coefficients determine the wing-normal and tangential forces, which are then transformed back into the physical $x$--$z$ frame.

A posture-dependent $2\times2$ added-mass matrix, computed using the boundary element method (BEM), accounts for the fluid inertia associated with translational acceleration. Together with the predicted aerodynamic forces and gravity, it determines the translational response of the agent. Time integration updates the flight velocity, position, and wing kinematics subject to prescribed motion constraints. The updated state and reward are returned to the learning algorithm, closing the feedback loop between wing motion, aerodynamic forces, and flight velocity. The aerodynamic formulation, BEM calculation, constraints, and training procedures are detailed in Methods.

\section{Results and discussion}\label{sec3}
\begin{figure}
\centering
\includegraphics[width=1.0\textwidth]{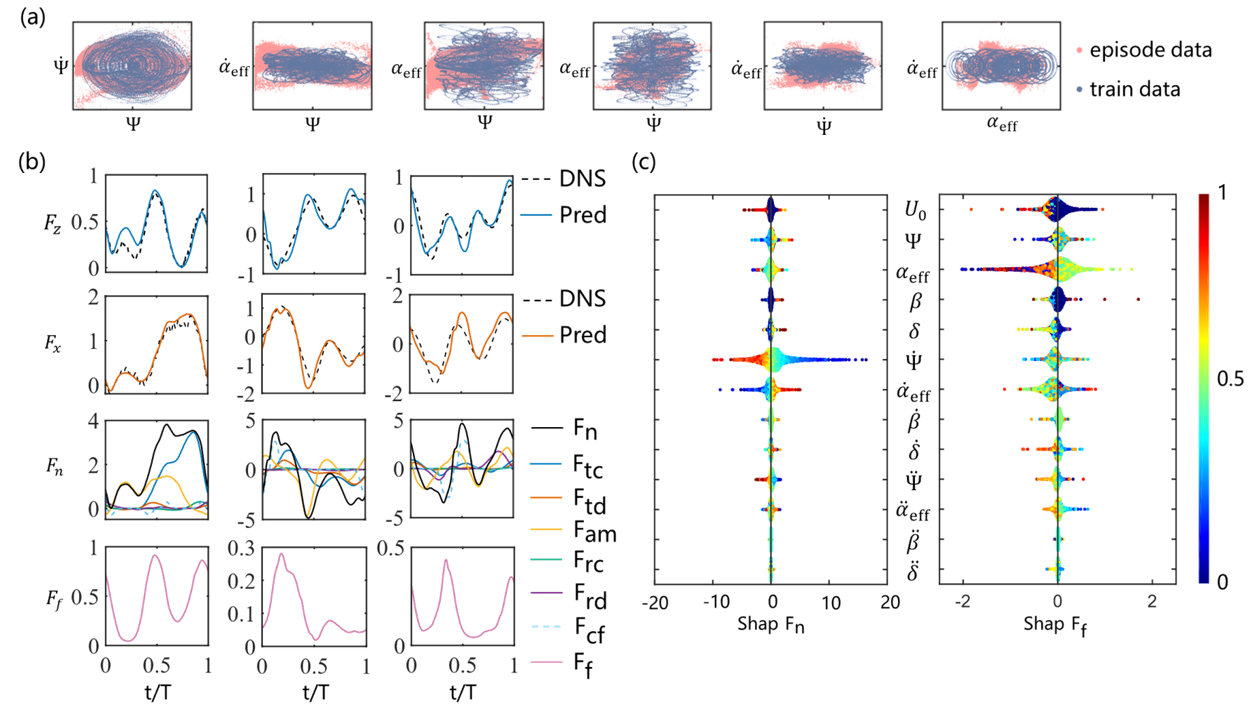}
\caption{Predictive performance and aerodynamic attribution of the multilayer perceptron for unsteady aerodynamic coefficients (AeroMLP).
(a) Pairwise projections of selected kinematic inputs comparing the DNS training data with states sampled from reinforcement-learning episodes.
(b) AeroMLP predictions and DNS results for three representative unseen test cases, together with the decomposition of the normal force $F_n$ and the tangential viscous force $F_f$.
(c) SHAP summary plots for $F_n$ and $F_f$, where color denotes the normalized value of each input variable.}\label{fig3}
\end{figure}
\subsection{AeroMLP captures unsteady butterfly aerodynamics}\label{3.1}
To assess whether AeroMLP operates within a data-supported domain during closed-loop flight, Fig.~\ref{fig3}(a) compares six pairwise projections of $\Psi$, $\dot{\Psi}$, $\alpha_{\mathrm{eff}}$, and $\dot{\alpha}_{\mathrm{eff}}$ between the DNS training data and 32,100 states collected from 214 reinforcement-learning episodes. The training data cover an average of $82.49\%$ of the agent states across these projections, while the mean displacement between their distribution centers is only $2.28\%$ of the corresponding training ranges. These results indicate that the randomly generated DNS training data capture most of the principal kinematic regimes explored by the agent, supporting the use of AeroMLP in closed-loop flight simulations.

Across an independent test set comprising 19 unseen cases, the prediction errors of AeroMLP for $F_x$ and $F_z$, normalized by the corresponding DNS force ranges, were $5.86\%$ and $4.84\%$, respectively. Figure~\ref{fig3}(b) shows that the model captures both the timing and magnitude of the principal force extrema across different conditions. The force decomposition further indicates that translational-circulation and added-mass forces dominate the normal force, consistent with the principal unsteady aerodynamic mechanisms identified in previous data-driven models and studies of fruit-fly flight.

By embedding the aerodynamic coefficients predicted by the multilayer perceptron into interpretable force expressions, AeroMLP learns the aerodynamic relationships represented by 66,734 instantaneous DNS training states. To determine how each input contributes to the model outputs, we use SHAP, which originates from the Shapley value in cooperative game theory. SHAP decomposes the difference between an individual model prediction and a reference prediction into additive feature contributions, whose signs and magnitudes describe the direction and strength of each contribution, respectively~\cite{shapley1953value,lundberg2017unified}. Figure~\ref{fig3}(c) presents the SHAP values for the wing-normal force $F_n$ and tangential viscous force $F_f$ predicted by AeroMLP, with color indicating the normalized value of each input variable. For $F_n$, $\dot{\Psi}$ has the strongest influence, while $\dot{\alpha}_{\mathrm{eff}}$ exhibits a clear directional association. Positive $\dot{\Psi}$, corresponding to wing elevation, is primarily associated with a negative contribution to $F_n$, whereas negative $\dot{\Psi}$ produces the opposite response. Positive $\dot{\alpha}_{\mathrm{eff}}$, corresponding to wing rotation in the pitch-up direction, is mainly associated with a positive contribution to $F_n$, whereas rotation in the opposite direction contributes negatively. This relationship is consistent with an important contribution of pitch-induced circulation to the wing-normal force, as suggested by experiments on rotating insect wings~\cite{dickinson1999rotation,sane2002rotation}. The attribution describes the surrogate, not an independent measurement of circulation. In comparison, $U_0$ has a less pronounced influence on $F_n$ because the wing-normal component of the incoming flow is jointly modulated by $\alpha_{\mathrm{eff}}$ and the wing-elevation angle $\Psi$. As $|\Psi|$ approaches $90^{\circ}$, the contribution of the incoming flow to the wing-normal velocity is strongly attenuated. For $F_f$, $\alpha_{\mathrm{eff}}$ is the most influential input. When $\alpha_{\mathrm{eff}}$ approaches $\pm90^{\circ}$, the tangential relative flow and viscous force both approach zero, leaving the aerodynamic load to be carried primarily by the wing-normal mechanisms. When $\alpha_{\mathrm{eff}}$ approaches $0^{\circ}$ or $\pm180^{\circ}$, the tangential flow velocity increases, together with the magnitude of the viscous force.

\begin{figure}
\centering
\includegraphics[width=1.0\textwidth]{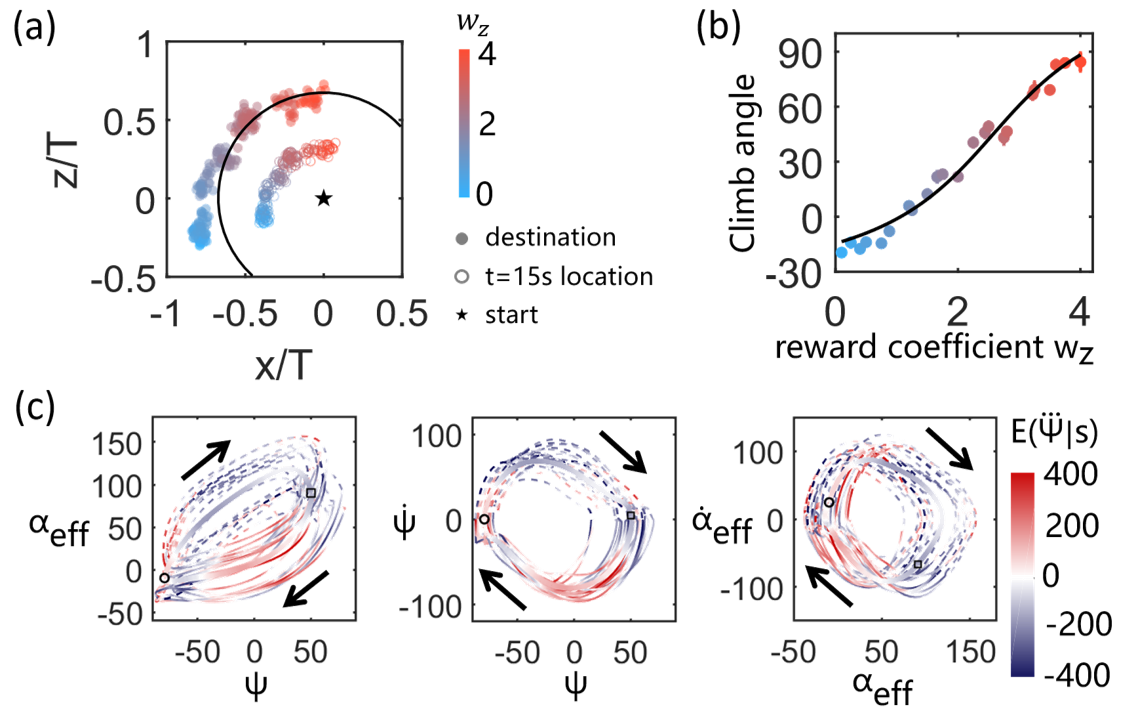}
\caption{Reward-directed climbing and the resulting wingbeat strategies. (a) Agent positions at mid-flight ($t=15~\mathrm{s}$, open circles) and at the end of each $30~\mathrm{s}$ episode (filled circles), coloured by the vertical-displacement reward coefficient $w_z$. The coordinates are normalized by the episode duration $T$, and the star denotes the initial position.
(b) Converged climbing angle as a function of $w_z$, with $w_x=4-w_z$. The solid curve represents the fitted reward--angle relationship.
(c) State-space phase portraits of the resulting wingbeats in the $(\psi,\alpha_{\mathrm{eff}})$, $(\psi,\dot{\psi})$, and $(\alpha_{\mathrm{eff}},\dot{\alpha}_{\mathrm{eff}})$ planes, coloured by the expected $\psi$-jerk output of the learned policy. Dashed and solid trajectories denote the upstroke and downstroke, respectively. The arrows indicate the
direction of temporal evolution along the phase trajectories.}
\label{fig4}
\end{figure}

\begin{figure}
\centering
\includegraphics[width=1.0\textwidth]{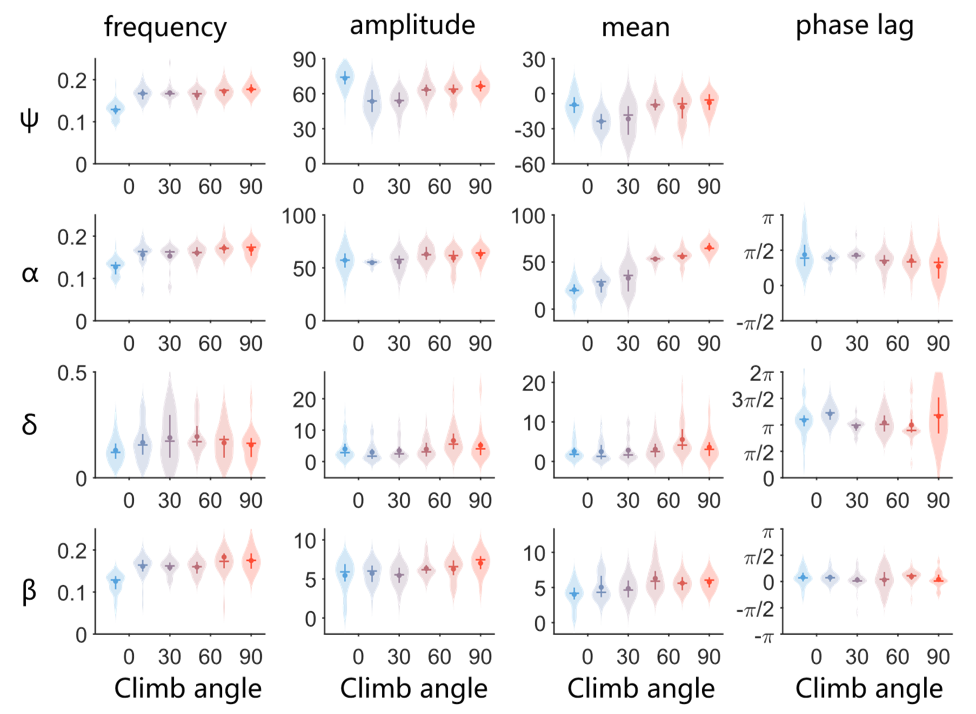}
\caption{Distributions of the fundamental frequency, amplitude, mean, and phase lag of the four wing kinematic angles at different climbing angles. Phase lags are measured relative to $\psi$.}
\label{fig4_d}
\end{figure}

\subsection{Self-generated wingbeats enable reward-directed climbing}\label{sec3.2}

To investigate how the reward configuration drives free climbing of the agent, we varied the relative weights of vertical and forward displacements in the reward function, with $w_x=4-w_z$. Figure~\ref{fig4}(a) shows that increasing $w_z$ continuously redirects the flight trajectory from forward descent to nearly vertical climbing. The endpoints obtained under different reward configurations approximately follow an arc, indicating that the reward primarily changes the direction of the net displacement while also affecting the mean flight speed. Over the same flight duration of $30~\mathrm{s}$, the mean net displacement of episodes with negative climbing angles is approximately $24.3\%$ larger than that of near-vertical climbing episodes. The converged mean climbing angle is described by
\begin{equation}
\gamma=
\tan^{-1}
\left(
\frac{w_z-1.062}{w_x+0.086}
\right),
\qquad w_x=4-w_z,
\end{equation}
with $R^2=0.975$, as shown in Fig.~\ref{fig4}(b). Thus, varying the reward weights continuously regulates the flight direction generated by the agent, rather than switching between a small number of discrete flight modes.

Figure~\ref{fig4}(c) further illustrates how the policy network uses $\psi$-jerk to organize the wingbeat motion. The colour represents the expected $\dddot{\psi}$ given by the policy distribution at each state. The $(\psi,\alpha_{\mathrm{eff}})$ phase portrait shows a stable phase relationship between the effective angle of attack and the elevation angle. The effective angle of attack increases during the upstroke and decreases during the downstroke. During the upstroke (dashed curves), the expected $\psi$-jerk is predominantly negative and reaches its negative extremum near mid-stroke. During the downstroke (solid curves), the expected $\psi$-jerk is predominantly positive and reaches its positive extremum near mid-stroke. The $(\psi,\dot{\psi})$ and $(\alpha_{\mathrm{eff}},\dot{\alpha}_{\mathrm{eff}})$ phase portraits further show that both wing elevation and pitching form stable closed trajectories. For ideal harmonic motion, an angle and its angular velocity trace an elliptical phase portrait with an exact phase difference of $\pi/2$. Although the present trajectories retain an approximately elliptical structure, their pronounced asymmetry and local deformation indicate a stable near-quadrature phase relationship together with wingbeat waveforms that depart from standard sinusoidal motion.

Figure~\ref{fig4_d} compares the fundamental waveform characteristics of the four kinematic angles across different climbing angles. The most pronounced variation occurs in the mean value of $\alpha$, which is strongly positively correlated with the climbing angle (Pearson $r=0.957$). On average, a $1^{\circ}$ increase in climbing angle is accompanied by an increase of approximately $0.65^{\circ}$ in the mean $\alpha$. The fundamental frequencies of $\psi$ and $\alpha$ also increase moderately with climbing angle, with correlation coefficients of $0.550$ and $0.598$, respectively. In contrast, the amplitude and mean of $\psi$, the amplitude of $\alpha$, and most waveform characteristics of the slot angle $\beta$ and sweep angle $\delta$ exhibit no clear monotonic dependence on climbing angle. Both $\delta$ and $\beta$ maintain stable phase lags relative to the elevation motion, with no clear monotonic trend across the sampled climbing angles. These results indicate that the agent primarily regulates its climbing direction by shifting the mean angle of attack and moderately increasing the wingbeat frequency, while the remaining kinematic parameters retain broadly similar distributions.

\subsection{Gravity projections shape climbing wingbeats}\label{sec4}

\begin{figure}
\centering
\includegraphics[width=1.0\textwidth]{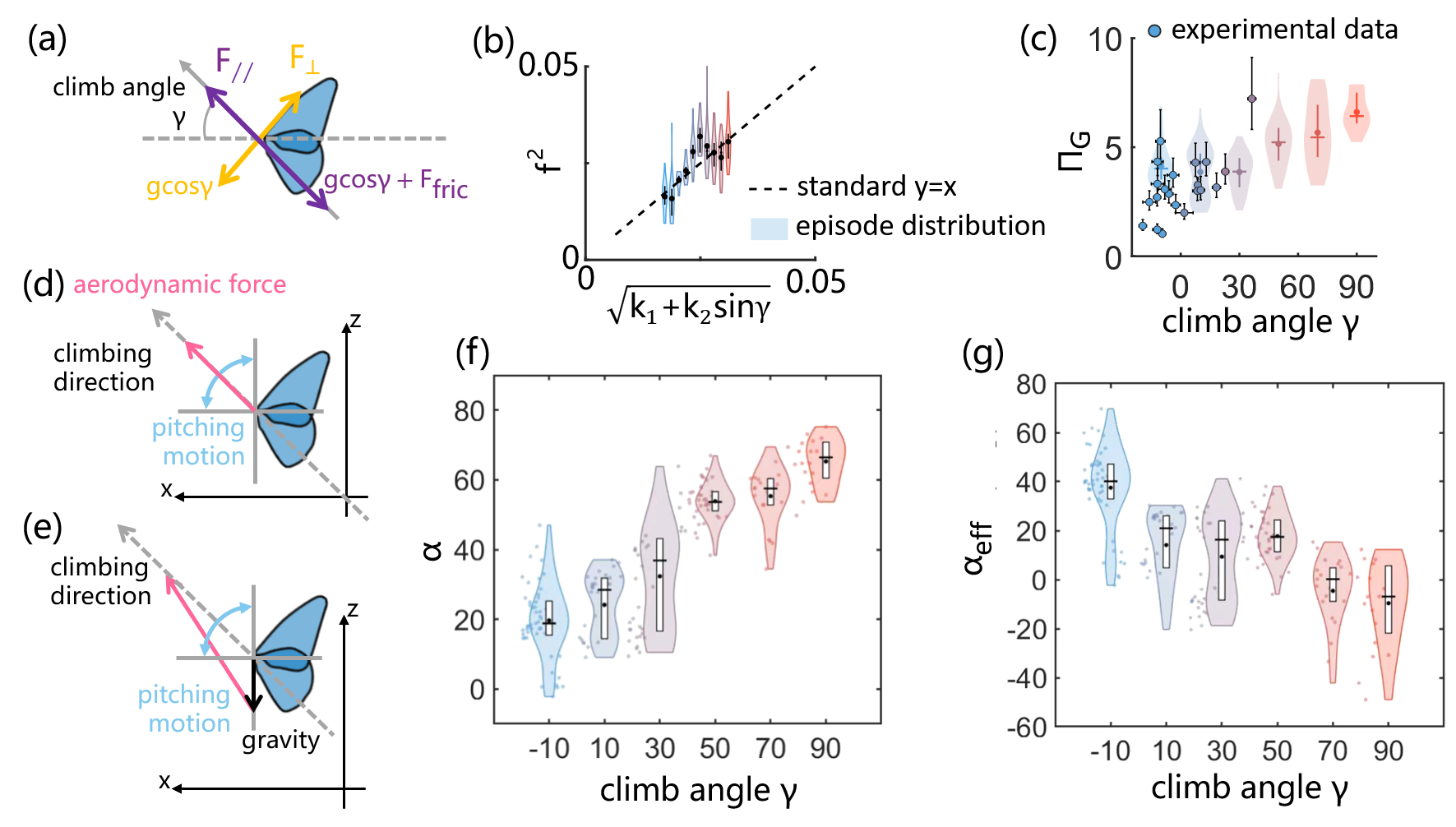}
\caption{Gravity-dependent force balance drives adaptations in climbing wingbeats. (a) Schematic force balance between the aerodynamic-force components and the projections of gravity at a climbing angle $\gamma$.
(b) Comparison between the squared wingbeat frequency obtained from the episodes and that predicted by the proposed scaling law. 
(c) Dynamic-pressure-to-weight ratio $\Pi_G$ as a function of climbing angle for experimentally observed butterflies and the virtual butterfly agent.
(d,e) Directions of the time-averaged pressure force required to sustain a climbing angle $\gamma$ in the absence and presence of gravity, respectively.
(f,g) Distributions of the time-averaged geometric angle of attack $\alpha$ and effective angle of attack $\alpha_{\mathrm{eff}}$, respectively, across different climbing angles.}
\label{fig5}
\end{figure}

Figure~\ref{fig5} further explains why the wingbeat frequency and mean geometric angle of attack vary systematically with climbing angle. Among the kinematic variables examined in Fig.~\ref{fig4_d}, these two quantities show the strongest positive correlations with climbing angle. Both changes represent dynamical responses of the agent to the varying projections of gravity along and normal to the flight direction. Figure~\ref{fig5} therefore uses the time-averaged force balance to explain the increase in wingbeat frequency and the physical relationship between angle of attack and climbing angle.

We distinguish the pressure force $\mathbf{F}_p$ from the total aerodynamic force $\mathbf{F}_{\mathrm a}=\mathbf{F}_p+\mathbf{F}_{\mathrm{fric}}$. Force direction matters because a wing-generated load can contribute to both propulsion and weight support~\cite{chin2019birds,johansson2021butterflies}. Consider a nearly straight segment with negligible cycle-averaged acceleration. Resolve the forces along the mean flight direction and its upward normal, and approximate the mean friction force as opposing flight with magnitude $D$. Writing $P=|\overline{\mathbf{F}}_p|$ gives
\begin{equation}
    \overline{F}_{p,\parallel}\simeq mg\sin\gamma+D,\qquad
    \overline{F}_{p,\perp}\simeq mg\cos\gamma,\qquad
    P^2\simeq (mg)^2+D^2+2mgD\sin\gamma .
\end{equation}
The amplitude varies little across the sampled climbs, so the characteristic wingtip speed scales with frequency. If the effective pressure coefficient and waveform shape also vary weakly, the pressure-force scale varies as $f^2$. Such velocity-based estimates are consistent with scaling approaches for oscillating foils~\cite{floryan2017scaling}, although the dependence on climbing angle below follows from our force balance. Treating $D$ and the proportionality factor as approximately constant gives
\begin{equation}
f^{2}\sim\sqrt{k_{1}+k_{2}\sin\gamma}.
\end{equation}
The episode data in Fig.~\ref{fig5}(b) generally follow this scaling relation, suggesting that the increase in wingbeat frequency primarily compensates for the growing projection of gravity along the climbing direction. Figure~\ref{fig5}(c) further shows that the dynamic-pressure-to-weight ratio of the agent increases with climbing angle and remains within a dynamical range comparable to that of experimentally observed butterflies.

Gravity also changes the required direction of the mean pressure force. Without gravity this force balances the assumed streamwise resistance. With gravity it must additionally supply an upward normal component. If $\chi$ denotes its inclination above the flight direction, then $P\sin\chi\simeq mg\cos\gamma$. To compare this force angle with the kinematics, we use the additional approximation $\overline{\alpha}_{\mathrm{eff}}\simeq\chi$ on the forward-propulsive branch. This is a model closure, not a geometric identity for an oscillating wing. With $c=P/(mg)$ treated as approximately constant for this angular fit, the resulting estimates are
\begin{equation}
\overline{\alpha}_{\mathrm{eff}}
\simeq
\sin^{-1}\left(\frac{\cos\gamma}{c}\right),
\qquad
\overline{\alpha}
\simeq
\gamma+\sin^{-1}\left(\frac{\cos\gamma}{c}\right),
\end{equation}
Fitting the episode data gives $c=2.39$. As the climbing angle increases, the normal projection of gravity decreases, leading to a reduction in $\overline{\alpha}_{\mathrm{eff}}$. The increase in geometric angle of attack is therefore jointly determined by the rotation of the climbing direction and the decrease in effective angle of attack, rather than following a one-to-one proportional relationship with $\gamma$. These expressions capture the first-order trends of the two mean angles shown in Fig.~\ref{fig5}(f,g), although appreciable variations remain among individual episodes.

\subsection{Frequency--pitch coupling redirects climbing trajectories}
\label{sec5}

\begin{figure}
\centering
\includegraphics[width=1.0\textwidth]{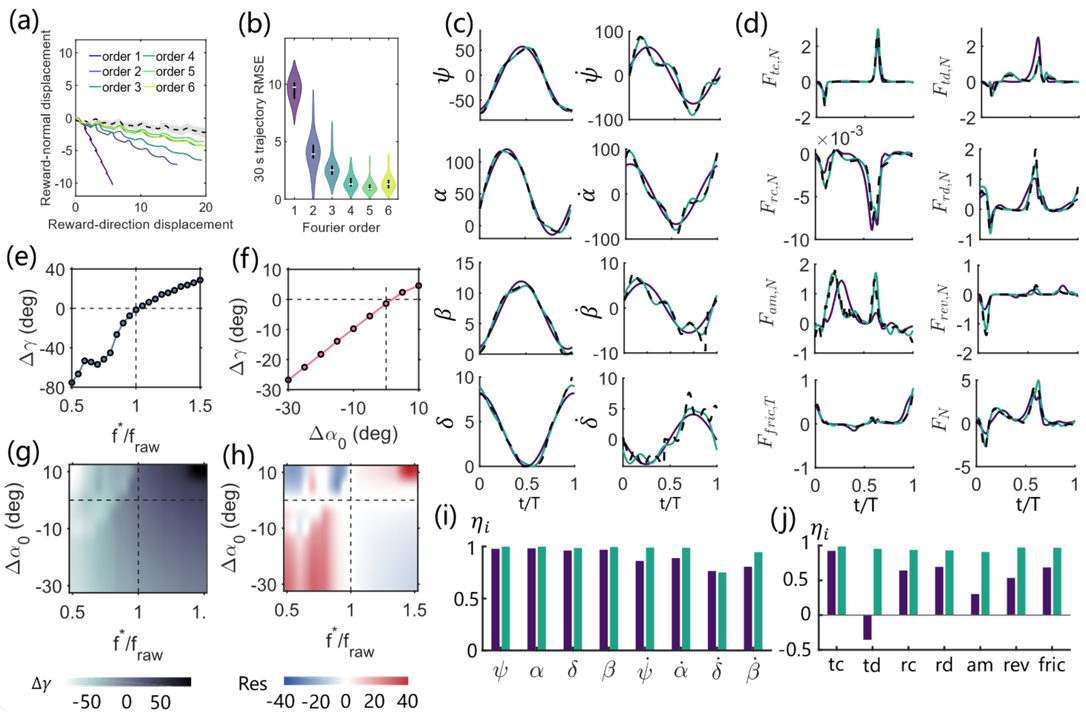}
\caption{Harmonic reconstruction of wing kinematics and the coupled effects of frequency and mean angle of attack on climbing. (a,b) Flight trajectories and their 30\,s trajectory RMSE relative to the original trajectories, obtained by fitting the four kinematic angles over a stable wingbeat cycle using first- to sixth-order harmonics. (c) Angles (left) and angular velocities (right) reconstructed using first-order (purple) and fourth-order (green) harmonics. (d) Aerodynamic-force components over one cycle for the two reconstructed motions. (e,f) Climbing-angle changes induced by varying frequency at fixed mean angle of attack and varying mean angle of attack at fixed frequency, respectively. (g,h) Climbing-angle changes and nonlinear interaction residuals under simultaneous modulation of both parameters. (i,j) Coefficients of determination $R^2$ of the reconstructed kinematic variables and aerodynamic-force components relative to the original data.}
\label{fig6}
\end{figure}

The preceding results identify the wingbeat frequency $f$ and mean geometric angle of attack $\overline{\alpha}$ as the two kinematic variables most strongly associated with climbing angle. These associations are descriptive rather than causal~\cite{altman2015association}. To test whether they can directly regulate climbing, we replace the original wing motion with a truncated Fourier series, allowing $f$ and $\overline{\alpha}$ to be varied independently while preserving the principal waveform characteristics. An order-$n$ reconstruction contains harmonics from $f$ to $nf$. Figure~\ref{fig6}(a) compares the resulting 30-s trajectories with the original episodes in a reward-aligned coordinate system. Across 20 high-reward episodes, the mean trajectory RMSE decreases from 9.53 for the first-order reconstruction to 1.48 for the fourth-order reconstruction, corresponding to a reduction of $84.47\%$. The errors for the fifth- and sixth-order reconstructions are 1.14 and 1.42, respectively, showing no consistent improvement beyond the fourth order [Fig.~\ref{fig6}(b)]. The fourth-order representation is therefore used for the subsequent modulation of frequency and mean angle of attack.

The large trajectory error of the first-order reconstruction arises from its inability to recover the amplitudes and phases of the angular velocities. Figure~\ref{fig6}(c,i) shows that the first-order reconstruction already reproduces the four kinematic angles with coefficients of determination of $R^2=0.96$--$0.98$, whereas the corresponding values for the angular velocities decrease to $0.76$--$0.89$. The fourth-order reconstruction increases the $R^2$ values of $\dot{\psi}$, $\dot{\alpha}$, and $\dot{\beta}$ to $0.95$--$0.99$, although the value for $\dot{\delta}$ remains approximately 0.75. These derivative quantities directly enter the aerodynamic formulation. The translational-circulation and pressure-drag forces depend on the wing-relative velocity governed by $\dot{\psi}$, the rotational terms depend on $\dot{\alpha}_{\mathrm{eff}}$, and the added-mass force further depends on angular acceleration. Consequently, the $R^2$ values of the aerodynamic components produced by the first-order reconstruction range from $-0.35$ to 0.92, including $-0.35$ for $F_{td,N}$ and 0.30 for $F_{am,N}$ [Fig.~\ref{fig6}(d,j)]. A negative $R^2$ indicates a reconstruction error larger than that obtained using the temporal mean. The fourth-order reconstruction increases the $R^2$ values of all aerodynamic components to $0.91$--$0.99$. Experiments on non-sinusoidal foil motions likewise identify peak lateral velocity as a useful predictor of propulsion~\cite{vanburen2017gaits}. Small differences in the reconstructed angles are therefore amplified through their time derivatives into substantial instantaneous force differences, which accumulate into different flight trajectories.

Using the fourth-order reconstruction, increasing the frequency from $0.5f_{\mathrm{raw}}$ to $1.5f_{\mathrm{raw}}$ produces an overall increase in the 30-s climbing angle from $-55.88^{\circ}$ to $48.23^{\circ}$ while $\overline{\alpha}$ is fixed. At fixed frequency, changing the mean angle of attack from $30^{\circ}$ below to $10^{\circ}$ above its original value increases the climbing angle from $-7.30^{\circ}$ to $24.06^{\circ}$. However, the combined response in Fig.~\ref{fig6}(g,h) cannot be obtained by linearly adding the two individual effects. We quantify this nonlinearity using the interaction residual
\begin{equation}
I(r_f,\Delta\overline{\alpha})
=
\gamma(r_f,\Delta\overline{\alpha})
-\gamma(r_f,0)
-\gamma(1,\Delta\overline{\alpha})
+\gamma(1,0),
\qquad
r_f=\frac{f}{f_{\mathrm{raw}}}.
\end{equation}
This residual contains changes in vertical displacement and is also affected by changes in forward displacement. At low frequency and reduced $\overline{\alpha}$, the agent undergoes a forward-gliding descent. The retained forward displacement slows the decline in climbing angle relative to reducing either frequency or $\overline{\alpha}$ alone, resulting in a positive $I$. At low frequency and increased $\overline{\alpha}$, the aerodynamic force remains insufficient while forward propulsion is weakened, accelerating the decline in climbing angle and producing a negative $I$. At high frequency and reduced $\overline{\alpha}$, a larger fraction of the aerodynamic force is directed forward, weakening the climbing benefit of the increased frequency. Only the combination of high frequency and increased $\overline{\alpha}$ provides sufficient aerodynamic-force magnitude while redirecting it vertically, producing a positive climbing synergy with a maximum interaction residual of $46.06^{\circ}$. Experiments on combined heaving and pitching also show that propulsion depends on the coordination of the two motions~\cite{vanburen2019scaling}. In the present model, frequency changes the velocity and acceleration scales of the wings, whereas the mean angle of attack changes both the aerodynamic coefficients and the force projection. Thus, $f$ and $\overline{\alpha}$ provide two low-dimensional controls of climbing direction, but their effects are not independent or linearly additive.

\section{Conclusion}\label{sec6}

To understand how aerodynamic forces shape butterfly climbing strategies, we developed a virtual framework for freely exploring wingbeat control. A DNS-informed AeroMLP was coupled with butterfly dynamics to construct an interactive aerodynamic environment. Within this environment, a soft actor--critic agent directly controlled the angular jerks of four wing kinematic angles without a prescribed wingbeat generator. The agent autonomously formed non-sinusoidal wingbeats and continuously redirected its flight from forward descent to near-vertical climbing, mainly by adjusting the wingbeat frequency and mean geometric angle of attack. Time-averaged force balance showed that these adaptations arise from changes in the projections of gravity along and normal to the climbing direction. Harmonic reconstruction and direct parameter modulation further demonstrated that frequency and mean angle of attack regulate climbing through nonlinear coupling rather than independent linear effects. This framework provides a route for uncovering adaptive flight strategies and developing freely controlled bio-inspired flapping-wing vehicles.

\section*{Methods}\label{sec11}

\subsection*{Agent state and wing-motion control}

The butterfly geometry is simplified from the morphology reported by Ancel et al.~\cite{ancel2017aerodynamic}. The model has four wing-kinematic degrees of freedom, represented by $\mathbf{q}=[\Psi,\alpha,\beta,\delta]^{\mathrm T}$. The center of mass translates in the $x$--$z$ plane. Body rotation is not included as an independent degree of freedom. The state supplied to the policy comprises 14 variables,
\begin{equation}
    \mathbf{s}
    =[U_x,U_z,\mathbf{q}^{\mathrm T},
    \dot{\mathbf{q}}^{\mathrm T},
    \ddot{\mathbf{q}}^{\mathrm T}]^{\mathrm T}.
\end{equation}
The four normalized policy outputs $u_i\in[0,1]$ are converted into angular jerks according to $j_i=(2u_i-1)j_{i,\max}$. Angular acceleration, angular velocity, and angle are advanced in time using these actions. Bounds on the wing angles and their time derivatives restrict the admissible motion. The reference scales used to define these bounds do not impose an oscillation frequency or a periodic waveform.

\subsection*{DNS data and aerodynamic-model training}

The aerodynamic dataset consists of instantaneous forces obtained from DNS under prescribed wing kinematics and flight velocities. Kinematic parameters are sampled across the conditions used to construct the training database. Complete simulation cases are assigned to the training, validation, and test sets, so that samples from the same case do not appear in different subsets. The independent test set is used only to evaluate the trained model.

Each aerodynamic input contains the equivalent flight speed, four wing angles, and their first and second time derivatives. Physical normalization is followed by feature-wise min--max scaling,
\begin{equation}
    \widehat{x}_i
    =\frac{x_i^{\mathrm p}-m_i}{M_i-m_i},
\end{equation}
where $x_i^{\mathrm p}$ denotes the physically normalized variable, and $m_i$ and $M_i$ are fixed reference bounds. The same transformations are used during training and deployment. Dataset inputs already stored in physically normalized form are not physically normalized a second time.

AeroMLP predicts aerodynamic coefficients rather than force components directly. The coefficients are inserted into a quasi-steady formulation based on Cai et al.~\cite{cai2021cfd}. The wing-normal force is represented as
\begin{equation}
    F_N=F_{\mathrm{tc},N}+F_{\mathrm{td},N}
    +F_{\mathrm{rc},N}+F_{\mathrm{rd},N}
    +F_{\mathrm{am},N}+F_{\mathrm{cf},N},
\end{equation}
where the terms represent translational circulation, translational drag, rotational circulation, rotational drag, added mass associated with wing motion, and clap-and-fling effects. Including a stroke-reversal contribution follows earlier quasi-steady treatments of fling circulation~\cite{armanini2016quasi}. Free-flight observations also identify clap-and-fling among several mechanisms used by butterflies~\cite{srygley2002unconventional,johansson2021butterflies}. A separate tangential contribution represents viscous friction. State-dependent coefficients allow these contributions to vary with the instantaneous flight condition.

The predicted forces are compared with the DNS forces during supervised training. Gradients are propagated through the force expressions to optimize the coefficient-prediction network. No reference labels for individual coefficients or force contributions are required. Accordingly, the force decomposition is a model-based interpretation, rather than an independently validated decomposition of the DNS forces. The trained aerodynamic model is held fixed during reinforcement learning.

\subsection*{Equivalent-flight representation}

Before aerodynamic evaluation, the translational velocity is converted into an equivalent speed and flight-path direction,
\begin{equation}
    U_{\mathrm{eq}}=\sqrt{U_x^2+U_z^2+\varepsilon_U},
    \qquad
    \theta=\operatorname{atan2}(U_z,U_x+\varepsilon_x).
\end{equation}
The regularization parameters prevent numerical singularities near zero velocity. In the coordinate convention used here, equivalent forward flight is directed along negative $x_0$. The effective angle of attack is therefore
\begin{equation}
    \alpha_{\mathrm{eff}}
    =\operatorname{atan2}\!\left[
    \sin(\alpha-\pi+\theta),\cos(\alpha-\pi+\theta)\right].
\end{equation}
The aerodynamic input replaces the two physical velocity components with the equivalent forward velocity and replaces the geometric angle of attack with $\alpha_{\mathrm{eff}}$, giving 13 input variables. In the present implementation, the rotational-rate inputs retain $\dot{\alpha}$ and $\ddot{\alpha}$ without adding derivatives of the flight-path angle. They therefore describe wing pitching rather than the exact time derivatives of $\alpha_{\mathrm{eff}}$.

The predicted normal and tangential forces are resolved into the equivalent flight plane. The resulting components are transformed back into the physical frame as
\begin{equation}
    \begin{bmatrix}F_x^{\mathrm a}\\F_z^{\mathrm a}\end{bmatrix}
    =
    \begin{bmatrix}
    -\cos\theta & \sin\theta\\
    -\sin\theta & -\cos\theta
    \end{bmatrix}
    \begin{bmatrix}F_{x,0}\\F_{z,0}\end{bmatrix}.
\end{equation}

\subsection*{Translational dynamics and added mass}

The aerodynamic response to wing motion is distinguished from the fluid inertia associated with acceleration of the translating model. For the latter, the wing surfaces are discretized into triangular elements, and a potential-flow boundary-element calculation gives a posture-dependent $2\times2$ added-mass matrix $\mathbf{M}_{\mathrm{add}}$ for translation in the $x$ and $z$ directions. The translational equation is
\begin{equation}
    \left(M\mathbf{I}_2+\mathbf{M}_{\mathrm{add}}\right)
    \dot{\mathbf{U}}
    =\mathbf{F}_{\mathrm a}-Mg\mathbf{e}_z.
\end{equation}
Here, $\mathbf{U}=[U_x,U_z]^{\mathrm T}$ and $\mathbf{e}_z$ is the upward unit vector. This approximation includes no term involving the time derivative of the added-mass matrix. Time integration advances the translational velocity, position, and wing state, after which the prescribed motion constraints are applied.

The added-mass matrix is obtained by solving two auxiliary potential-flow problems at each instantaneous wing configuration. The geometry is held fixed during each calculation. The surrounding fluid is assumed to be incompressible, inviscid, and irrotational, with velocity $\mathbf{u}=\nabla\phi$. For unit translation in direction $k\in\{x,z\}$, the potential $\phi^{(k)}$ satisfies
\begin{equation}
    \nabla^2\phi^{(k)}=0
    \quad\text{in }\Omega_f,\qquad
    \frac{\partial\phi^{(k)}}{\partial n}=n_k
    \quad\text{on }S,\qquad
    \phi^{(k)}\rightarrow0
    \quad\text{as }|\mathbf{x}|\rightarrow\infty,
\end{equation}
where $\Omega_f$ is the exterior fluid domain, $S$ is the closed surface of the finite-thickness wing geometry, and $\mathbf{n}$ points from the solid into the fluid. The surface boundary condition enforces no penetration.

Using the free-space Green function
$G(\mathbf{x},\mathbf{y})=(4\pi|\mathbf{x}-\mathbf{y}|)^{-1}$,
Green's identity gives the exterior boundary integral equation
\begin{equation}
    \frac{1}{2}\phi^{(k)}(\mathbf{x})
    -\operatorname{p.v.}\int_S
    \frac{\partial G(\mathbf{x},\mathbf{y})}{\partial n_{\mathbf{y}}}
    \phi^{(k)}(\mathbf{y})\,\mathrm{d}S_{\mathbf{y}}
    =
    -\int_S G(\mathbf{x},\mathbf{y})n_k(\mathbf{y})\,\mathrm{d}S_{\mathbf{y}},
\end{equation}
where $\operatorname{p.v.}$ denotes the principal value. The coefficient $1/2$ applies at a smooth surface point.

The surface is divided into $N$ planar triangular elements. The potential is approximated as constant on each element, and the integral equation is collocated at the element centroids $\mathbf{x}_i$. Defining
\begin{equation}
    H_{ij}=\int_{S_j}G(\mathbf{x}_i,\mathbf{y})\,\mathrm{d}S_{\mathbf{y}},
    \qquad
    D_{ij}=\operatorname{p.v.}\int_{S_j}
    \frac{\partial G(\mathbf{x}_i,\mathbf{y})}{\partial n_{\mathbf{y}}}
    \,\mathrm{d}S_{\mathbf{y}},
\end{equation}
the two translation problems are assembled as
\begin{equation}
    \mathbf{A}\boldsymbol{\Phi}=-\mathbf{H}\mathbf{N},
    \qquad
    \mathbf{A}=\frac{1}{2}\mathbf{I}-\mathbf{D},
    \qquad
    \boldsymbol{\Phi}
    =[\boldsymbol{\phi}^{(x)},\boldsymbol{\phi}^{(z)}],
    \qquad
    N_{ik}=n_{i,k}.
\end{equation}
Regular element integrals are evaluated by numerical quadrature, with refined quadrature for nearby elements. The weakly singular self integral $H_{ii}$ is evaluated using a singularity-resolving quadrature. For a planar element collocated at its centroid, $D_{ii}=0$ and hence $A_{ii}=1/2$. The matrix $\mathbf{A}$ is factorized once for each configuration, and the same factorization is used to solve both right-hand sides.

The added-mass coefficients follow from the surface potentials,
\begin{equation}
    [\mathbf{M}_{\mathrm{add}}]_{\ell k}
    =-\rho\int_S\phi^{(k)}n_\ell\,\mathrm{d}S
    \approx
    -\rho\sum_{i=1}^{N}
    \phi_i^{(k)}n_{i,\ell}\Delta S_i,
    \qquad \ell,k\in\{x,z\}.
\end{equation}
Equivalently, these coefficients satisfy
\begin{equation}
    [\mathbf{M}_{\mathrm{add}}]_{\ell k}
    =\rho\int_{\Omega_f}
    \nabla\phi^{(\ell)}\cdot\nabla\phi^{(k)}\,\mathrm{d}V,
\end{equation}
which establishes the symmetry and positive semidefiniteness of the continuous added-mass matrix. The diagonal entries describe translational fluid inertia, while the off-diagonal entries describe coupling between the $x$ and $z$ directions. This frozen-configuration calculation supplies the instantaneous acceleration-dependent contribution used in the translational equation. Terms involving the time derivative of the added-mass matrix are not included.

\subsection*{Reinforcement-learning procedure}

A stochastic policy is trained using SAC to maximize displacement in a prescribed direction. The displacement reward combines vertical and forward progress,
\begin{equation}
    r_t=c_r\left(
    w_z\Delta z_t-w_x\Delta x_t-\lambda_p P_t
    \right),
    \qquad w_z+w_x=4,
\end{equation}
where forward motion corresponds to decreasing $x$, $P_t$ is a nonnegative constraint penalty, and $c_r$ scales the reward. Separate policies are trained for different reward weights. No reference wing trajectory or wingbeat generator is supplied.

At each interaction step, the policy samples an action from the normalized state. The environment updates the wing motion and translational dynamics and returns the next state, reward, and termination flag. These transitions are stored in a replay buffer and sampled for SAC updates. Each episode spans 3,000 steps with a time increment of $0.01$, corresponding to a simulated duration of $30$. The reward weights specify the desired displacement preference rather than directly prescribing the resulting climbing angle.

Algorithm~\ref{alg:training} summarizes the training procedure. The aerodynamic surrogate remains fixed during reinforcement learning, while the policy learns wing-motion control through repeated interaction with the virtual environment.

\begin{algorithm}[t]
\caption{Learning wingbeat control through aerodynamic feedback}
\label{alg:training}
\begin{algorithmic}[1]
\Require Trained AeroMLP, reward weights $(w_x,w_z)$,
         time step $\Delta t$, motion constraints
\Ensure Trained policy $\pi_\theta$
\State Initialize SAC networks and replay buffer $\mathcal{D}$
\For{each training episode}
    \State Reset position, translational velocity, and wing state
    \State Encode the initial state $\widehat{\mathbf{s}}_0$
    \For{$t=0,\ldots,T_{\max}-1$}
        \State Sample $\mathbf{u}_t
               \sim\pi_\theta(\cdot\mid\widehat{\mathbf{s}}_t)$
        \State Convert actions to angular jerks:
               $j_{i,t}=(2u_{i,t}-1)j_{i,\max}$
        \State Construct the equivalent-flight aerodynamic input
        \State Predict aerodynamic coefficients using AeroMLP
        \State Evaluate normal and tangential aerodynamic forces
        \State Transform the forces into the physical $x$--$z$ frame
        \State Evaluate $\mathbf{M}_{\mathrm{add}}$
               at the current wing configuration
        \State Compute translational acceleration from
               aerodynamic forces, gravity, and effective inertia
        \State Advance wing kinematics, flight velocity, and position
               over $\Delta t$, applying the motion constraints
        \State Compute $r_t$ from displacement and constraint penalties
        \State Encode $\widehat{\mathbf{s}}_{t+1}$
               and determine termination flag $d_t$
        \State Store $(\widehat{\mathbf{s}}_t,\mathbf{u}_t,r_t,
               \widehat{\mathbf{s}}_{t+1},d_t)$ in $\mathcal{D}$
        \If{sufficient transitions are available}
            \State Sample a minibatch from $\mathcal{D}$
            \State Update SAC critics, policy, and entropy temperature
            \State Soft-update the target networks
        \EndIf
        \If{$d_t$}
            \State \textbf{break}
        \EndIf
    \EndFor
\EndFor
\end{algorithmic}
\end{algorithm}

\backmatter

\bmhead{Acknowledgements}

This work is supported by NSFC the Excellence Research Group Program for multiscale problems in nonlinear mechanics (Grant No. 12588201), the National Natural Science Foundation of China (Grant No. 12425207), the Chinese Academy of Sciences Project for Young Scientists in Basic Research (Grant No. YSBR-087), the Strategic Priority Research Program of Chinese Academy of Sciences (Grant No. XDB0620102), the National Natural Science Foundation of China (Grant No. 52501409) and the Postdoctoral Fellowship Program of CPSF (Grant No. GZB20240773).

\section*{Declarations}

\bmhead{Competing interests}
The authors declare no competing interests.

\end{document}